\documentclass[twocolumn]{aastex631}
\usepackage{CJK}
\usepackage{subfigure}
\received{March 1, 2021}
\revised{April 1, 2021}
\accepted{\today}
\published{\today}

\submitjournal{\colorbox{yellow}{\bf\color{blue}ApJ}}
\submitjournal{\bf\color{blue}ApJ}
\submitjournal{ApJ}

\shorttitle{Hunting for Changing-Look Blazar candidates}
\shortauthors{Kang et al.}
\graphicspath{{./}{figures/}}
\begin{document}
\begin{CJK*}{UTF8}{gbsn}
%
%

\title[Hunting for Changing-Look Blazar candidates]
{Hunting for the candidates of Changing-Look Blazar using $Mclust$ Clustering Analysis}

\correspondingauthor{Shi-Ju Kang}
\email{kangshiju@alumni.hust.edu.cn}

\author[0000-0002-9071-5469]{Shi-Ju Kang}
\affiliation{School of Physics and Electrical Engineering,  Liupanshui Normal University,  Liupanshui, Guizhou, 553004, People's Republic of China}

\author[0009-0003-0317-4831]{Shan-Shan Ren}

\affiliation{Institute of Space Sciences, Shandong University, Weihai, Shandong, 264209, People's Republic of China}

\author[0000-0003-0170-9065]{Yong-Gang Zheng}
\affiliation{Department of Physics, Yunnan Normal University, Kunming, Yunnan, 650092, People's Republic of China}

\author[0000-0003-4773-4987]{Qingwen Wu}
\affiliation{Department of Astronomy, School of Physics, 
Huazhong University of Science and Technology, Wuhan, Hubei, 430074, People's Republic of China}



\begin{abstract}

The changing-look blazars (CLBs) are the blazars that their optical spectral lines at different epochs show a significant changes and present a clear transition between the standard FSRQ and BL Lac types. The changing-look phenomena in blazars are highly significant for enhancing our understanding of certain physical problems of active galactic nuclei (AGNs), such as the potential mechanism of the state transition in the accretion process of the supermassive black holes in the central engine of AGNs, the possible intrinsic variation of the jet, and the connection between the accretion disk and the jet. Currently, the CLBs reported in the literature are still rare astronomical objects. In our previous work, we found that there are 8 physical properties parameters of CLBs located between those of FSRQs and those of BL Lacs. In order to search more CLB candidates (CLBCs), we employed the $mclust$ Gaussian Mixture Modelling clustering algorithm to perform clustering analysis for the 255 subsets of the 8 physical properties parameters with 2250 blazars from the 4FGL-DR3. We find that there are 29 subsets with 3 groups (corresponding to bl lacs, fsrqs, and CLBCs), in which there are 4 subsets with the adjusted Rand index greater then 0.610 (ARI $>$ 0.610). The combined clustering results from 4 subsets report that there are 111 CLBCs that includes 44 CLBs reported in previous literature and 67 new CLBCs, where 11 CLBCs labeled as BL Lac and 56  CLBCs labeled as FSRQ in 4FGL catalog.

\end{abstract}

\keywords{ Active galactic nuclei (16) -- Blazars (164) --- BL Lacertae objects (158) --- Flat-spectrum radio quasars (2163) }



\section{Introduction} \label{sec:intro}

Blazars are an extreme subclass of radio-loud active galactic nuclei (AGNs), whose relativistic jets along the observer's line of sight point to  the Earth \citep{1995PASP..107..803U}. Based on the strength of the optical spectral lines (e.g., equivalent width, EW, of the spectral line is greater or less than 5 \AA), blazars come in two flavors: flat spectrum radio quasars (FSRQs) with strong EW (EW $\ge$ 5 \AA), and BL Lacerate objects (BL Lacs) that the spectral lines are fainter (EW $<$ 5 \AA) or even absent in their optical spectra \citep[e.g.,][]{1991ApJ...374..431S,1991ApJS...76..813S}.

\vspace{2.5mm} 

The Changing-Look Blazars (CLBs) are enhancement/appearance or a diminution/disappearance of optical spectral lines at different epochs  (e.g., \citealt{2016AJ....151...32A};   \citealt{2021ApJ...913..146M}; \citealt{2021AJ....161..196P};  \citealt{2021Univ....7..372F,2022Univ....8..587F}；\citealt{2023MNRAS.525.3201K}; \citealt{2024ApJ...962..122K}, and references therein) and present a clear transition (e.g., EW become larger or smaller) between the standard FSRQ and BL Lac types. 

These changing-look phenomena in blazars are important (of great significance) to shed light on understanding of the physical reasons for the presence of CLB sources,  the divergent properties of BL Lacs and FSRQs (e.g., \citealt{2014ApJ...797...19R}; \citealt{2021AJ....161..196P}; \citealt{2024ApJ...962..122K}), the possible intrinsic variation and the radiation mechanism of the blazar jets (e.g., \citealt{2012MNRAS.420.2899G}; \citealt{2021ApJ...913..146M}), and the possible mechanism of state transition of the accretion process in supermassive black holes of the central engine of AGNs, and the accretion disk-jet connection, and the black hole-galaxy co-evolution (e.g., \citealt{2014ApJ...797...19R}; \citealt{2021AAS...23740807M}).

\vspace{2.0mm}

At present, the mechanism of the transition between FSRQ and BL Lac states is still unclear. These rare phenomenons are commonly addressed by some possible scenarios in the previous literature (e.g., see \citealt{2014ApJ...797...19R}; \citealt{2021AJ....161..196P}; \citealt{2021ApJ...913..146M} for the related discussions and references therein). For instance, 1) the broad lines (EW) of CLBs may be swamped by the strong  (beamed) jet continuum variability (e.g., \citealt{1995ApJ...452L...5V}; \citealt{2012MNRAS.420.2899G}; \citealt{2014ApJ...797...19R}; \citealt{2019RNAAS...3...92P}). 2) The broad lines may be also diluted by strong jet continuum variability caused by the jet bulk Lorentz factor variability (e.g., \citealt{2009A&A...496..423B}). 3) The broad lines may be also overwhelmed by the non-thermal continuum for these transition sources with weak radiative cooling (e.g., \citealt{2012MNRAS.425.1371G,2023arXiv231005096P}). 4) Furthermore,  some broad lines of the FSRQs may be not observed due to with a high redshift (e.g., $z > 0.7$, \citealt{2015MNRAS.449.3517D}), where the one of the strongest $H{\alpha}$ line may fall outside the optical window. In addation, 5) several observational effects (e.g., signal-to-noise ratio, spectral resolution, etc.) may also affect the optical classification (see \citealt{2021AJ....161..196P} for the related discussions). In addition to jet effects and observational effects, several explanations for the origin of the changing-look AGNs (CLAGNs) are also used to discuss the origin of CLBs or to argue the transition mechanism between FSRQs and BL Lacs. Similar to  CLAGNs, 6) the CLBs may also be argued that originate from the variable obscuration, such as the dusty clouds moving into/out of our line-of-sight and show a significant variation (e.g., \citealt{2019A&A...625A..54H,2022arXiv221105132R}), or 7) a sudden change in accretion rate that resulted in the broad emission lines become stronger or weaker (e.g., \citealt{2022ApJ...936..146X} and \citealt{2024arXiv240217099R} for the related discussions and references therein).

\vspace{2.0mm}

Recently, as the research deepens, more and more CLBs have been discovered and reported. For instance，\cite{2021ApJ...913..146M} reported a CLB B2 1420+32 (also named OQ 334), which experienced the transitions between FSRQ and BL Lac states multiple times over a several years timescale. \cite{2021AJ....161..196P} reported 26 CLBs that changed their classification, where three of them are confident in the changing-look nature based on the optical spectra available in the Large Sky Area Multi-object Fibre Spectroscopic Telescope (LAMOST) Data Release 5 (DR5) archive. Based on the EW of some blazars, \cite{2022ApJ...936..146X} reported 52 CLBs, and declared 45 of them are newly confirmed CLBs. In \cite{2021Univ....7..372F,2022Univ....8..587F}, they compiled a gamma-ray jetted AGN sample based on the 4FGL catalog. They reported 34 CLAGNs, 32 of them are labeled as blazars (24 FSRQs, 7 BL Lacs, and 1 BCU) in 4FGL catalog, based on a significant change in optical spectral lines  (disappearance and reappearance) in different observation epochs reported in the previous literature (see \citealt{2021Univ....7..372F,2022Univ....8..587F} for more details and references therein). \cite{2024ApJ...968...81P} reports that the EW of the CIV 1550 emission line of PKS 0446+112 changes dramatically from 20 to 0.8. In addition to discovering some CLBs based on optical spectra, researchers also predicted some CLB candidates based on statistical analysis. For instance, in \cite{2022ApJ...935....4Z}, they reported 46 blazars that are likely to be candidates of CLBs based on the analysis of the broad line region luminosity in Eddington units.
\cite{2023MNRAS.525.3201K} suggested that there are 157 false BL Lacs that are possible intrinsically FSRQs misclassified as BL Lacs.
Comparing the CLBs reported in the literatures, we argue that these  false BL Lacs are the likely candidates of CLBs.

\vspace{2.0mm}
In our previous work, we found that there are 8 variables of CLBs ( $\Gamma_{\rm ph}$, $\alpha_{\rm ph}$, ${\rm HR}_{34}$,  ${\rm HR}_{45}$, CD, $L_{\rm disk}$, $\lambda$=$L_{\rm disk}/L_{\rm Edd}$, and $z$, see Section \ref{sec:sample}) with the density distributions for CLBs located between those of BL Lacs and those of FSRQs, based on the univariate analysis,  bivariate analysis, and multivariate analysis (\citealt{2024ApJ...962..122K}). These properties can be used to search more CLB candidates. In order to address the issue, the $mclust$ Gaussian Mixture Modelling clustering algorithm is  employed to perform clustering analysis, to search or evaluate more CLB candidates.
The sample and $mclust$ algorithms are introduced in Section 2.
The results are presented in Section 3.
The discussion and conclusion are shown in Section 4.



\section{Sample and Method} \label{sec:SampleMethod}

\subsection{Sample} \label{sec:sample}

We selected all 2250 blazars (1,397 BL Lacs, 105 CLBs, and 748 FSRQs) with the known classifications in the 4FGL-DR3 catalog under CLASS in the FITS\footnote{\url{https://fermi.gsfc.nasa.gov/ssc/data/access/lat/12yr_catalog/gll_psc_v31.fit}} table.
The gamma-ray photon index $\Gamma_{\rm ph}$ and the spectral slope ($\alpha_{\rm ph}$ at E0, photon index at Pivot Energy when fitting with are direct collected from LogParabola) are directly obtained from the 4FGL catalogs (4LAC-DR3, \citealt{2022ApJS..263...24A}, 4FGL-DR3, \citealt{2022ApJS..260...53A}), which compile 12 years of Fermi-LAT data.
The hardness ratios (e.g., see \citealt{2012ApJ...753...83A}) ${\rm HR}_{34}$ (3: 300MeV$-$1GeV; 4: 1$-$3GeV), and ${\rm HR}_{45}$ (4: 1$-$3GeV; 5: 3$-$10GeV) are calculated by the Equation, \(HR_{i j}=\frac{\nu F \nu_{j}-\nu F \nu_{i}}{\nu F \nu_{j}+\nu F \nu_{i}}\), in  \cite{2024ApJ...962..122K} based on the spectral energy distribution ($\nu{F}\nu$) collected from \cite{2022ApJS..260...53A}, where $i,j = $3, 4 and 5 are indices corresponding to the different spectral energy bands. Among of them, there are 1667 blazars (818 BL Lacs, 101 CLBs and 748 FSRQs) with redshift measurements (z) reported in \cite{2022ApJS..263...24A}; there are 925 blazars (307 BL Lacs, 95 CLBs and 523 FSRQs) with the measurements of the Compton dominance (CD; the ratio of the inverse Compton to synchrotron peak luminosities) and the luminosity of accretion disk ($L_{\rm disk}$) in the Eddington units ($\lambda = L_{\rm disk}/L_{\rm Edd}$) reported in \cite{2021ApJS..253...46P}. Where the 105 CLBs are obtained from an online changing-look (transition)  blazars catalog (TCLB Catalog, S.-J. Kang et al. 2024, in preparation) that archived on Zenodo\footnote{\url{https://www.zenodo.org/record/10061349}} (\citealt{shi_ju_kang_2023_10061349}, also see  \citealt{2024ApJ...962..122K} for the details).

\subsection{Method} \label{sec:Method}

\vspace{2.0mm}
Unsupervised learning utilizes algorithms for data exploration, such as clustering, dimensionality reduction, and outlier detection, to find patterns in unlabeled datasets without human guidance. This approach is valuable for scientific discovery, including in fields like astronomy where it can reveal novel insights (e.g., \citealt{2019arXiv190407248B}). Cluster analysis, a key unsupervised learning method, organizes data based on similarities or differences without prior knowledge, using various algorithms like exclusive, overlapping, hierarchical, and probabilistic clustering.

\vspace{2.0mm}
In this work, 
$mclust$ R package\footnote{\url{https://cran.r-project.org/web/packages/mclust/index.html}} (\citealt{mclust_book}) is employed to perform clustering analysis, as the selected parameters in the sample exhibit a predominantly or approximately normal distribution.
The $mclust$ is a popular R package complied by \cite{mclust_2016} using R Language \citep{R_code} for model-based clustering, classification, and density estimation based on finite Gaussian mixture modelling. This package provides several tools for model-based clustering, classification, and density estimation, which include Bayesian regularization and dimension reduction based on the Expectation-Maximization (EM) algorithm (e.g.,  \citealt{mclust_2016}; \citealt{mclust_book}).

The EM algorithm iteratively optimizes parameter estimation. It starts by initializing parameters and then computes expectations until convergence.
The ``ellipsoidal, equal, volume''(EVV) model is that each cluster has an ellipsoidal shape and that the volume of all clusters is equal.
The number of mixing components and the covariance parameterisation are selected using the Bayesian Information Criterion (BIC).

\vspace{2.0mm}

The adjusted Rand index (ARI; \citealt{Hubert_Arabie_1985}) is used for evaluating a clustering solution. The ARI is a measure of agreement between two partitions,  one estimated by a statistical procedure independent of the labelling of the groups, and one being the true classification. It has zero expected value in the case of a random partition, and it is bounded above by 1, with higher values representing better partition accuracy.

\vspace{2.0mm}

For the 8 physical properties parameters, there are  255 subsets (where the number of subsets of a set of n elements excluding the empty set).
Then using the $mclust$ software package for clustering analysis, first use the EM algorithm to estimate the parameters required by the EVV model.
Subsequently, according to the results of the EM algorithm, use the EVV model to describe the clustering structure of the data.
Then calculate the BIC value to evaluate the fitting effect of different models. After determining the best model and parameters, finally calculate the ARI value to evaluate the quality of the clustering results.
The 255 subsets are evaluated and listed in Table \ref{tab:parameter_subsets}.

\vspace{2.0mm}

In addition, NbClust Package for determining the best number of clusters (\citealt{NbClust2014}) is also employed to test the number of clusters that use the $mclust$ package.
Which provides 30 indices for determining the number of clusters and proposes to user the best clustering scheme from the different results obtained by varying all subsets of number of clusters, distance measures, and clustering methods (see \citealt{NbClust2014} for the details).


%


\vspace{2.0mm}
\begin{deluxetable*}{lccccccccccccccc}
\tablenum{1}
\tablecaption{parameter subsets \label{tab:parameter_subsets}}
\tablewidth{0pt}
\tablehead{
\colhead{No.} & \colhead{$N_{s}$} & \colhead{Group} & \colhead{BIC} &\colhead{$N_{\rm C}$} &\colhead{$N_{\rm B}$} &\colhead{$N_{\rm F}$} &
\colhead{ARI} &\colhead{par1} &\colhead{par2} &\colhead{par3} &\colhead{par4} &\colhead{par5} &\colhead{par6} &\colhead{par7} &\colhead{par8}
}
\decimalcolnumbers
\startdata
No.4&2250&3&-1088.87&34&148&2068&0.067&4&...&...&...&...&...&...&...\\
No.14&921&3&-2275.67&111&367&443&0.486&1&7&...&...&...&...&...&...\\
No.18&925&3&-1835.91&7&319&599&0.602&2&5&...&...&...&...&...&...\\
No.27&925&3&-1729.24&46&308&571&0.478&4&5&...&...&...&...&...&...\\
No.28&925&3&-2808.23&4&437&484&0.509&4&6&...&...&...&...&...&...\\
No.29&921&3&-2562.86&270&280&371&0.402&4&7&...&...&...&...&...&...\\
No.32&921&3&-3480.41&237&340&344&0.482&5&7&...&...&...&...&...&...\\
No.39&925&3&364.21&64&241&620&0.455&1&2&5&...&...&...&...&...\\
No.53&921&3&-3139.73&185&330&406&0.568&1&5&7&...&...&...&...&...\\
No.63&925&3&-1379.22&153&234&538&0.371&2&4&5&...&...&...&...&...\\
No.68*&921&3&-3538.33&161&316&444&0.628&2&5&7&...&...&...&...&...\\
No.84&921&3&-3438.6&279&298&344&0.443&4&5&7&...&...&...&...&...\\
No.86&921&3&-3954.52&200&337&384&0.394&4&6&7&...&...&...&...&...\\
No.89*&921&3&-4843.68&156&320&445&0.613&5&6&7&...&...&...&...&...\\
No.90&912&3&-3968.33&7&329&576&0.595&5&6&8&...&...&...&...&...\\
No.105&921&3&-1893.46&65&346&510&0.57&1&2&6&7&...&...&...&...\\
No.120&912&3&-1715.24&143&301&468&0.447&1&4&5&8&...&...&...&...\\
No.121&921&3&-3112.42&191&360&370&0.437&1&4&6&7&...&...&...&...\\
No.124*&921&3&-4492.58&165&306&450&0.625&1&5&6&7&...&...&...&...\\
No.144&921&3&-4908.69&189&307&425&0.587&2&5&6&7&...&...&...&...\\
No.158*&921&3&-4832.36&151&310&460&0.636&4&5&6&7&...&...&...&...\\
No.161&908&3&-4124.59&131&334&443&0.519&4&6&7&8&...&...&...&...\\
No.179&921&3&-2789.53&69&335&517&0.586&1&2&5&6&7&...&...&...\\
No.193&921&3&-3975.64&223&342&356&0.431&1&4&5&6&7&...&...&...\\
No.212&908&3&-4960.91&140&319&449&0.538&2&5&6&7&8&...&...&...\\
No.218&908&3&-4917.08&144&327&437&0.522&4&5&6&7&8&...&...&...\\
No.239&908&3&-4125.63&190&329&389&0.416&1&4&5&6&7&8&...&...\\
No.245&908&3&-4675.04&136&323&449&0.503&2&4&5&6&7&8&...&...\\
No.252&908&3&-2437.69&97&321&490&0.51&1&2&4&5&6&7&8&...\\
...&...&...&...&...&...&...&...&...&...&...&...&...&...&...&...\\
\enddata
\tablecomments{
The ordinal number  (No.)  of the parameter subsets is listed in Column 1.
The number ($N_{s}$)  of sources used in clustering analysis for corresponding to each subsets is listed in Column 2.
The BIC with the components (Groups) obtained that fitted by $mclust$ model  are listed in Column 4, and 3 respectively. 
The number of results for clustering are listed in Column 5, 6, and 7 respectively.
Where
$N_{\rm C}$ represent the CLBCs, 
$N_{\rm B}$ represent the bl lacs, 
$N_{\rm F}$ represent the fsrqs predicted in $mclust$  clustering analysis, respectively.
ARI is listed in Column 8.
The labels of the parameters are shown in Columns 9-16, where 
1: $\Gamma_{\rm ph}$;
2: $\alpha_{\rm ph}$;
3: ${\rm HR}_{34}$;
4: ${\rm HR}_{45}$;
5: CD;
6: $L_{\rm disk}$;
7: $\lambda$=$L_{\rm disk}/L_{\rm Edd}$;
8: $z$, respectively.
A portion parameter subsets (Groups = 3) are shown here for guidance regarding their form and content.
}
\end{deluxetable*}

\vspace{-15.0mm}
\begin{deluxetable*}{cccccccccccccc}
\tablenum{2}
\tablecaption{the predictions for the selected parameter subsets \label{tab:combinations_predictions}}
\tablewidth{0pt}
\tablehead{
\colhead{4FGL Name} & \colhead{Class} 
                                                                & \colhead{$\Gamma_{\rm ph}$} & \colhead{$\alpha_{\rm ph}$} 
                                                                &\colhead{ ${\rm HR}_{34}$} &\colhead{ ${\rm HR}_{45}$} 
                                                                &\colhead{CD} & \colhead{$L_{\rm disk}$} 
                                                                &\colhead{$\lambda$=$L_{\rm disk}/L_{\rm Edd}$} 
                                                                &\colhead{$z$} 
                                                                &\colhead{No.68} &\colhead{No.89} 
                                                                &\colhead{No.124} &\colhead{No.158} 
}
\decimalcolnumbers
\startdata
4FGL J1954.6$-$1122	&	CLB 	&	2.41 	&	2.37 	&	-0.25 	&	-0.23 	&	0.58 	&	44.46 	&	-1.70 	&	0.68 	&	CLBC	&	CLBC	&	CLBC	&	CLBC	\\
4FGL J1959.1$-$4247	&	FSRQ	&	2.44 	&	2.24 	&	-0.19 	&	-0.31 	&	0.63 	&	46.15 	&	-1.35 	&	2.17 	&	CLBC	&	fsrq    &	CLBC	&	fsrq    \\
4FGL J2000.9$-$1748	&	FSRQ	&	2.21 	&	2.17 	&	-0.12 	&	-0.08 	&	0.23 	&	44.91 	&	-1.41 	&	0.65 	&	CLBC	&	CLBC	&	CLBC	&	CLBC	\\
4FGL J2026.0$-$2845	&	CLBC 	&	2.59 	&	2.50 	&	-0.43 	&	-0.30 	&	0.39 	&	44.81 	&	-1.84 	&	0.88 	&	CLBC	&	CLBC	&	CLBC	&	CLBC	\\
4FGL J2032.0$+$1219	&	CLB 	&	2.46 	&	2.43 	&	-0.21 	&	-0.28 	&	0.22 	&	44.94 	&	-1.48 	&	1.22 	&	CLBC	&	CLBC	&	CLBC	&	CLBC	\\
4FGL J2035.4$+$1056	&	FSRQ	&	2.40 	&	2.32 	&	-0.17 	&	-0.36 	&	0.27 	&	45.23 	&	-0.91 	&	0.60 	&	fsrq    &	CLBC	&	fsrq    &   fsrq    \\
4FGL J2056.2$-$4714	&	FSRQ	&	2.45 	&	2.35 	&	-0.30 	&	-0.41 	&	0.63 	&	45.83 	&	-1.58 	&	1.49 	&	CLBC	&	CLBC	&	CLBC	&	CLBC	\\
4FGL J2115.4$+$2932	&	FSRQ	&	2.37 	&	2.32 	&	-0.22 	&	-0.24 	&	0.74 	&	45.63 	&	-1.42 	&	1.51 	&	fsrq    &	fsrq    &	CLBC	&	fsrq    \\
4FGL J2118.0$+$0019	&	FSRQ	&	2.52 	&	2.55 	&	-0.25 	&	-0.83 	&	0.19 	&	44.80 	&	-0.91 	&	0.46 	&	fsrq    &	CLBC	&	fsrq    &	fsrq    \\
4FGL J2120.6$-$6114	&	FSRQ	&	2.35 	&	2.29 	&	-0.11 	&	-0.23 	&	0.07 	&	45.07 	&	-1.20 	&	1.02 	&	CLBC	&	CLBC	&	CLBC	&	CLBC	\\
4FGL J2134.2$-$0154	&	CLB 	&	2.28 	&	2.21 	&	-0.16 	&	-0.14 	&	0.02 	&	44.83 	&	-2.03 	&	1.28 	&	CLBC	&	CLBC	&	CLBC	&	CLBC	\\
4FGL J2151.4$+$4156	&	BL Lac 	&	2.09 	&	2.04 	&	0.26 	&	-0.07 	&	-0.46 	&	44.55 	&	-0.80 	&	0.49 	&	CLBC	&	BL Lac	&	BL Lac	&	CLBC	\\
4FGL J2152.5$+$1737	&	CLB 	&	2.43 	&	2.41 	&	-0.34 	&	-0.22 	&	-0.18 	&	44.41 	&	-2.31 	&	0.87 	&	CLBC	&	BL Lac	&	BL Lac	&	BL Lac	\\
4FGL J2158.1$-$1501	&	CLB 	&	2.17 	&	2.07 	&	-0.06 	&	-0.22 	&	0.05 	&	44.86 	&	-1.36 	&	0.67 	&	CLBC	&	CLBC	&	CLBC	&	CLBC	\\
4FGL J2204.3$+$0438	&	CLB 	&	2.35 	&	2.40 	&	-0.10 	&	-0.48 	&	-0.09 	&	42.76 	&	-2.67 	&	0.03 	&	BL Lac	&	CLBC	&	CLBC	&	CLBC	\\
4FGL J2206.8$-$0032	&	CLB 	&	2.25 	&	2.16 	&	-0.08 	&	-0.09 	&	0.07 	&	44.64 	&	-2.05 	&	1.05 	&	CLBC	&	CLBC	&	CLBC	&	CLBC	\\
4FGL J2212.0$+$2356	&	CLB 	&	2.23 	&	2.15 	&	-0.11 	&	-0.22 	&	0.56 	&	45.62 	&	-1.18 	&	1.13 	&	CLBC	&	fsrq    &	fsrq    &	fsrq    \\
4FGL J2212.9$-$2526	&	FSRQ	&	2.25 	&	2.09 	&	-0.09 	&	-0.16 	&	-0.06 	&	46.67 	&	-0.19 	&	1.83 	&	CLBC	&	fsrq    &	fsrq    &	fsrq    \\
4FGL J2216.9$+$2421	&	CLBC 	&	2.25 	&	2.22 	&	-0.15 	&	-0.16 	&	-0.09 	&	45.04 	&	-1.51 	&	0.50 	&	CLBC	&	CLBC	&	CLBC	&	CLBC	\\
4FGL J2226.8$+$0051	&	FSRQ	&	2.49 	&	2.81 	&	-0.40 	&	-0.39 	&	0.38 	&	45.50 	&	-1.37 	&	2.26 	&	fsrq    &	CLBC	&	fsrq    &	CLBC	\\
4FGL J2236.3$+$2828	&	CLB 	&	2.23 	&	2.12 	&	-0.08 	&	-0.26 	&	0.26 	&	45.65 	&	-0.90 	&	0.79 	&	CLBC	&	fsrq    &	fsrq    &	fsrq    \\
4FGL J2237.0$-$3921	&	FSRQ	&	2.42 	&	2.37 	&	-0.20 	&	-0.29 	&	0.08 	&	44.83 	&	-1.07 	&	0.30 	&	CLBC	&	CLBC	&	CLBC	&	CLBC	\\
4FGL J2243.7$-$1231	&	BL Lac 	&	2.13 	&	2.04 	&	0.04 	&	-0.17 	&	-0.06 	&	44.51 	&	-2.07 	&	0.77 	&	CLBC	&	BL Lac	&	CLBC	&	CLBC	\\
4FGL J2244.2$+$4057	&	CLB 	&	2.13 	&	2.05 	&	-0.02 	&	-0.10 	&	0.40 	&	45.30 	&	-1.27 	&	1.17 	&	CLBC	&	CLBC	&	CLBC	&	CLBC	\\
4FGL J2247.5$-$3700	&	FSRQ	&	2.46 	&	2.47 	&	-0.28 	&	0.15 	&	0.17 	&	46.69 	&	0.29 	&	2.25 	&	fsrq    &	fsrq    &	CLBC	&	CLBC	\\
4FGL J2250.7$-$2806	&	BL Lac 	&	2.63 	&	2.64 	&	-0.05 	&	-0.56 	&	1.13 	&	43.84 	&	-2.59 	&	0.34 	&	CLBC	&	CLBC	&	CLBC	&	CLBC	\\
4FGL J2315.6$-$5018	&	CLB 	&	2.38 	&	2.37 	&	-0.24 	&	-0.03 	&	-0.32 	&	44.71 	&	-1.81 	&	0.81 	&	CLBC	&	BL Lac	&	BL Lac	&	BL Lac	\\
4FGL J2321.9$+$3204	&	FSRQ	&	2.25 	&	2.14 	&	-0.07 	&	-0.22 	&	0.46 	&	45.71 	&	-1.18 	&	1.49 	&	CLBC	&	fsrq    &	CLBC	&	fsrq    \\
4FGL J2323.5$-$0317	&	FSRQ	&	2.26 	&	2.15 	&	-0.15 	&	-0.25 	&	0.53 	&	47.09 	&	-0.31 	&	1.39 	&	fsrq    &	fsrq    &	CLBC	&	fsrq    \\
4FGL J2332.1$-$4118	&	FSRQ	&	2.27 	&	2.13 	&	0.12 	&	-0.37 	&	0.01 	&	46.24 	&	-0.69 	&	0.67 	&	CLBC	&	fsrq    &	fsrq    &	fsrq    \\
4FGL J2338.1$+$0325	&	FSRQ	&	2.46 	&	2.45 	&	-0.26 	&	-0.34 	&	0.04 	&	43.82 	&	-2.16 	&	0.27 	&	CLBC	&	CLBC	&	CLBC	&	CLBC	\\
4FGL J2345.2$-$1555	&	CLB 	&	2.16 	&	2.04 	&	-0.04 	&	-0.14 	&	0.30 	&	45.32 	&	-1.30 	&	0.62 	&	CLBC	&	CLBC	&	CLBC	&	CLBC	\\
4FGL J2348.0$-$1630	&	FSRQ	&	2.30 	&	2.24 	&	-0.14 	&	-0.23 	&	0.21 	&	45.32 	&	-1.23 	&	0.58 	&	CLBC	&	CLBC	&	CLBC	&	CLBC	\\
4FGL J2349.2$+$4535	&	BL Lac 	&	1.81 	&	1.74 	&	0.24 	&	0.08 	&	-0.16 	&	42.85 	&	-2.97 	&	0.04 	&	BL Lac	&	BL Lac	&	BL Lac	&	CLBC	\\
4FGL J2349.4$+$0534	&	CLB 	&	2.44 	&	2.42 	&	0.11 	&	-0.38 	&	-0.10 	&	44.08 	&	-2.06 	&	0.42 	&	CLBC	&	BL Lac	&	CLBC	&	CLBC	\\
4FGL J2354.6$+$4554	&	FSRQ	&	2.32 	&	2.03 	&	-0.04 	&	-0.42 	&	0.56 	&	46.13 	&	-1.18 	&	1.99 	&	CLBC	&	fsrq    &	CLBC	&	fsrq    \\
4FGL J2354.9$+$8151	&	FSRQ	&	2.50 	&	2.34 	&	-0.22 	&	-0.45 	&	0.06 	&	45.12 	&	-1.82 	&	1.34 	&	CLBC	&	CLBC	&	CLBC	&	CLBC	\\
4FGL J2357.4$-$0152	&	CLB 	&	2.28 	&	2.08 	&	0.21 	&	-0.39 	&	0.21 	&	43.91 	&	-1.53 	&	0.81 	&	CLBC	&	CLBC	&	CLBC	&	CLBC	\\
4FGL J2358.5$-$1808	&	BL Lac 	&	2.06 	&	2.02 	&	-0.18 	&	0.03 	&	0.01 	&	43.27 	&	-2.23 	&	0.20 	&	CLBC	&	CLBC	&	CLBC	&	CLBC	\\
$...$               &    $...$	&  $...$    &	$...$   &	$...$  	&	$...$  	&	$...$ 	&	$...$  	&	$...$  	&	$...$ 	& $...$     & $...$     &  $...$    &  $...$    \\
\enddata
\tablecomments{
The 4FGL names are presented in Column 1.
The optical classes reported in 4FGL or CLBs reported in the previous literature are presented in  Column 2.
Columns 3-10 shows the 
$\Gamma_{\rm ph}$; $\alpha_{\rm ph}$; ${\rm HR}_{34}$; ${\rm HR}_{45}$;
CD; $L_{\rm disk}$; $\lambda$=$L_{\rm disk}/L_{\rm Edd}$ and $z$, respectively.
Columns 11-14 shows the results of the predictions for  
No.68, No.89, No.124, and No.158 subsets respectively.
A portion results are shown here for guidance regarding their form and content. (This table is available in its entirety in machine-readable form.)}
\end{deluxetable*}

\begin{figure*}[htp!]
\centering
\includegraphics[width=18cm,height=10cm]{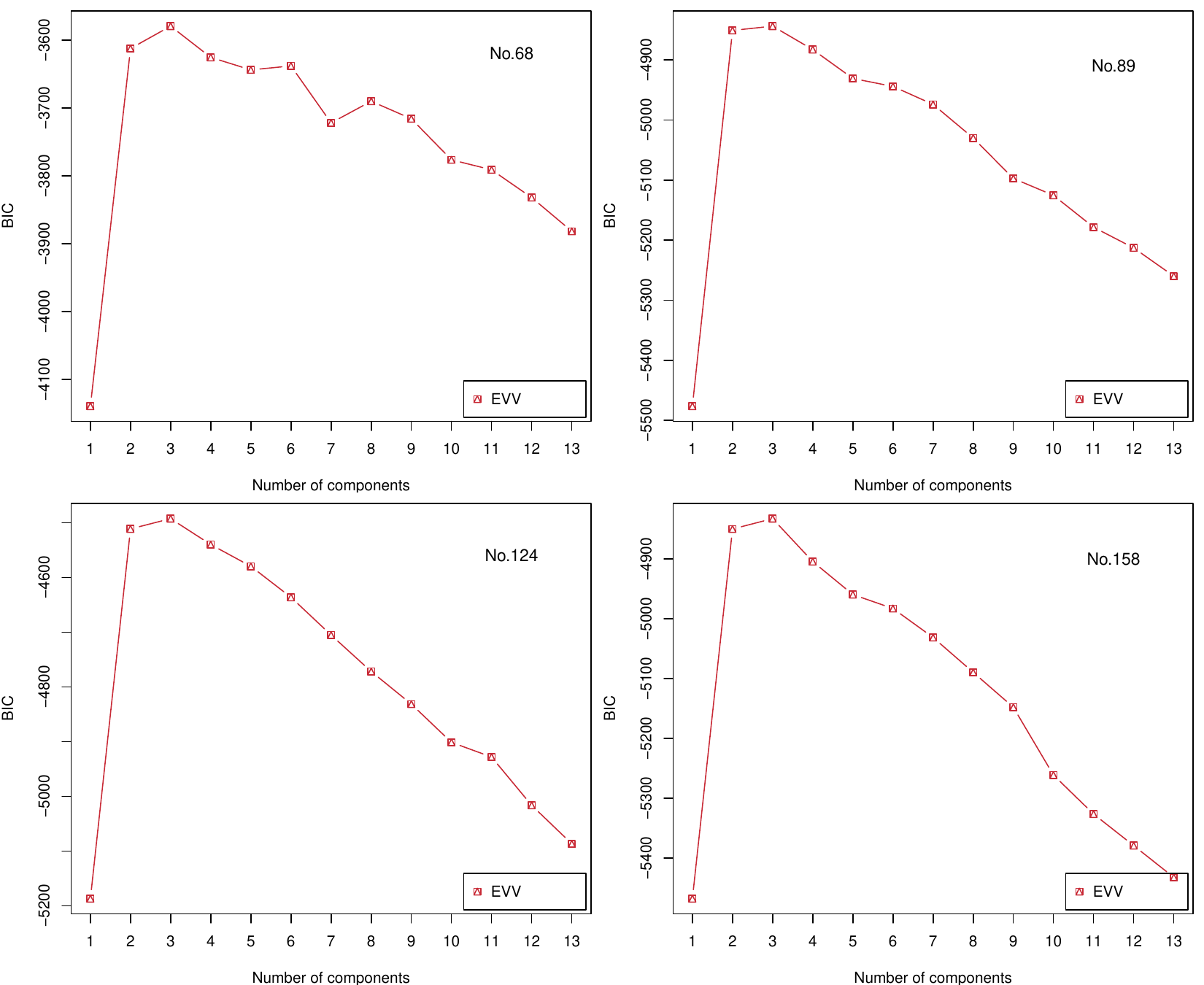}
\caption{The BIC in the $mclust$ clustering analysis for the No.68, No.89, No.124, and No.158 subsets respectively. 
\label{Fig_BIC}}
\end{figure*}

\begin{figure}[htp!]
\centering
\includegraphics[width=8cm,height=5cm]{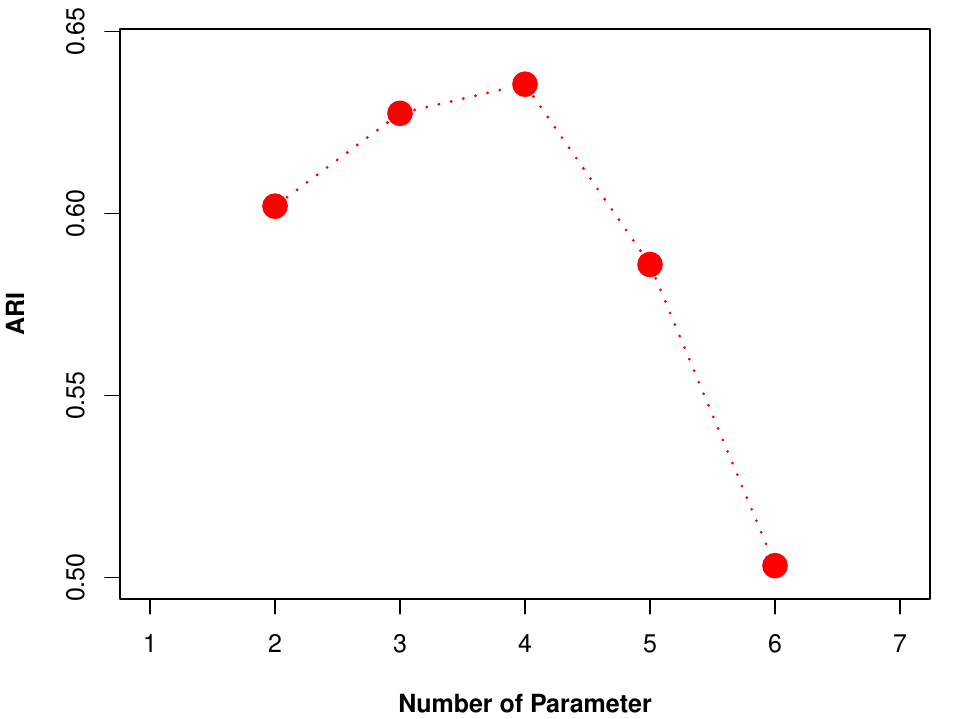}	
\caption{The ARI for different subsets of parameters with 3 groups  (fsrqs, bl lacs, and CLBCs) in the $mclust$ clustering analysis.
\label{Fig_ARI}}
\end{figure}

\begin{figure*}[htp!]
\centering
\includegraphics[width=18cm,height=10cm]{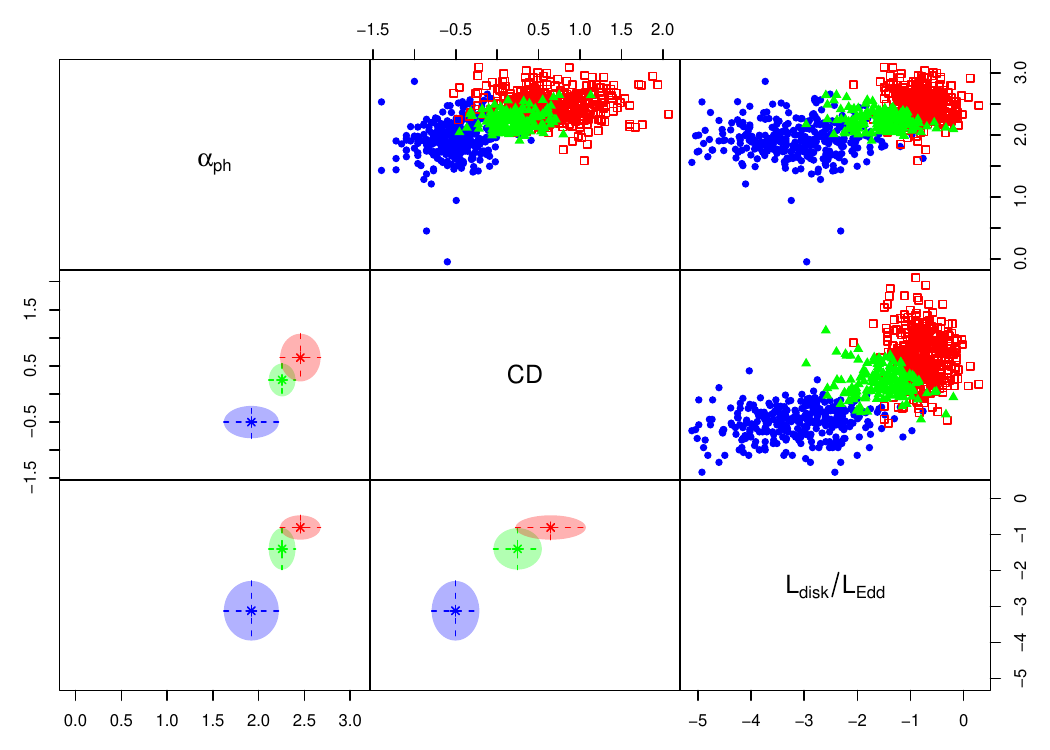}   
\includegraphics[width=18cm,height=10cm]{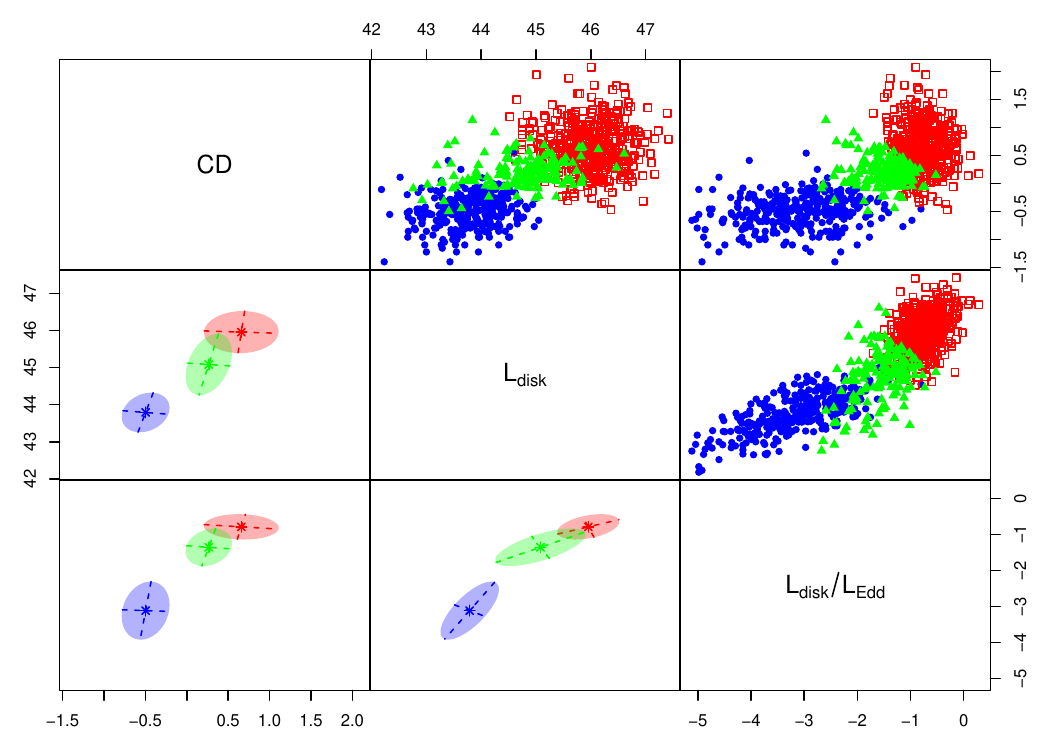}		
\caption{
The Scatterplots of the prediction results in clustering analysis for 
the No.68 subsets ($\alpha_{\rm ph}$, CD, and $\lambda$=$L_{\rm disk}/L_{\rm Edd}$) (top panels)
and the No.89 subsets (CD, $L_{\rm disk}$, and $\lambda$=$L_{\rm disk}/L_{\rm Edd}$) (low pannels),
where the red hollow squares, and blue solid dots, and green triangles 
indicate the predictions: fsrqs, bl lacs, and CLBCs respectively.
The Gaussian fitting results of each pairwise parameters are shown in the bottom left panels (the mirroring of upper right panels).
\label{Fig_combinations_3par}}
\end{figure*}

\begin{figure*}[htp!]
\centering
\includegraphics[width=18cm,height=10cm]{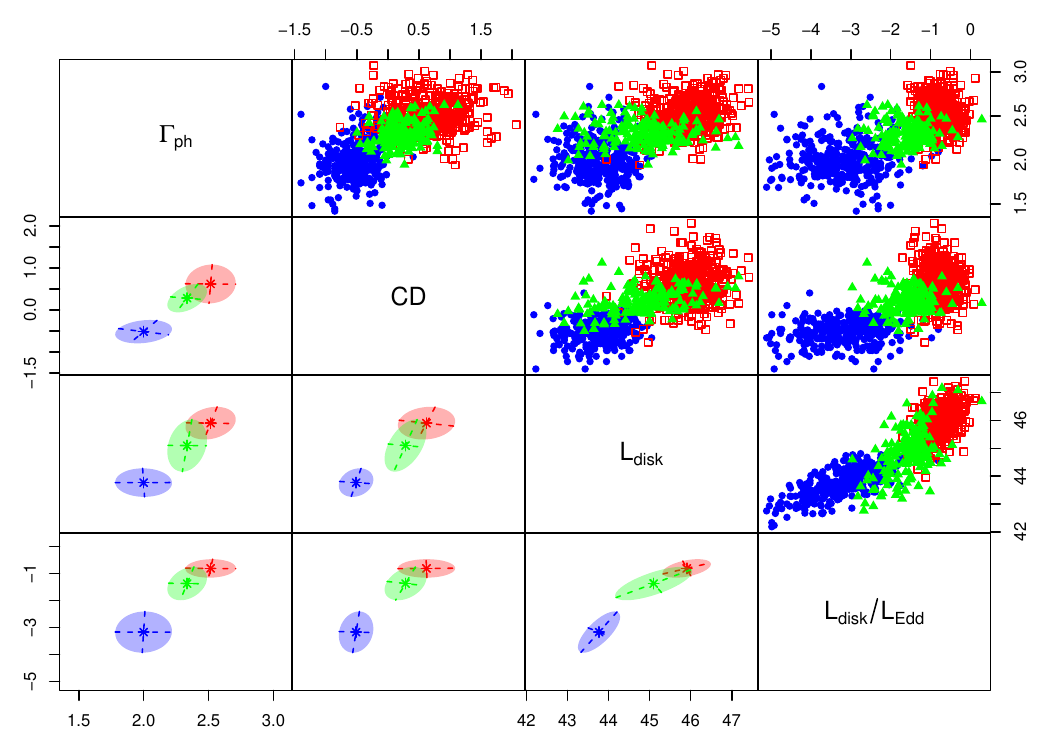}
\includegraphics[width=18cm,height=10cm]{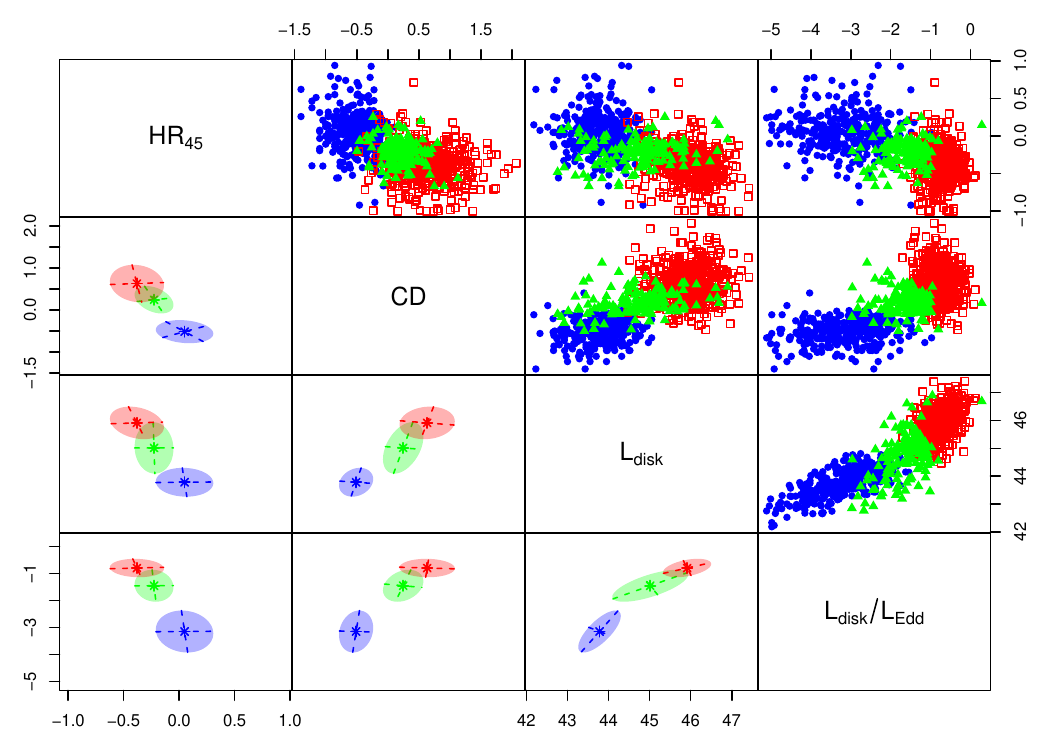}		
\caption{
The Scatterplots of the prediction results in clustering analysis for 
the No.124 subsets  ($\Gamma_{\rm ph}$, CD, $L_{\rm disk}$, and  $\lambda$=$L_{\rm disk}/L_{\rm Edd}$) (top panels)
and the No.158 subsets (${\rm HR}_{45}$, CD, $L_{\rm disk}$, and  $\lambda$=$L_{\rm disk}/L_{\rm Edd}$) (low panels),
where the red hollow squares, and blue solid dots, and green triangles 
indicate the predictions: fsrqs, bl lacs, and CLBCs respectively. The Gaussian fitting results of each pairwise parameters are shown in the bottom left panels (the mirroring of upper right panels).
\label{Fig_combinations_4par}}
\end{figure*}

\begin{figure*}[htp!]
\centering
 \vspace{2.0mm}
\includegraphics[width=18cm,height=18cm]{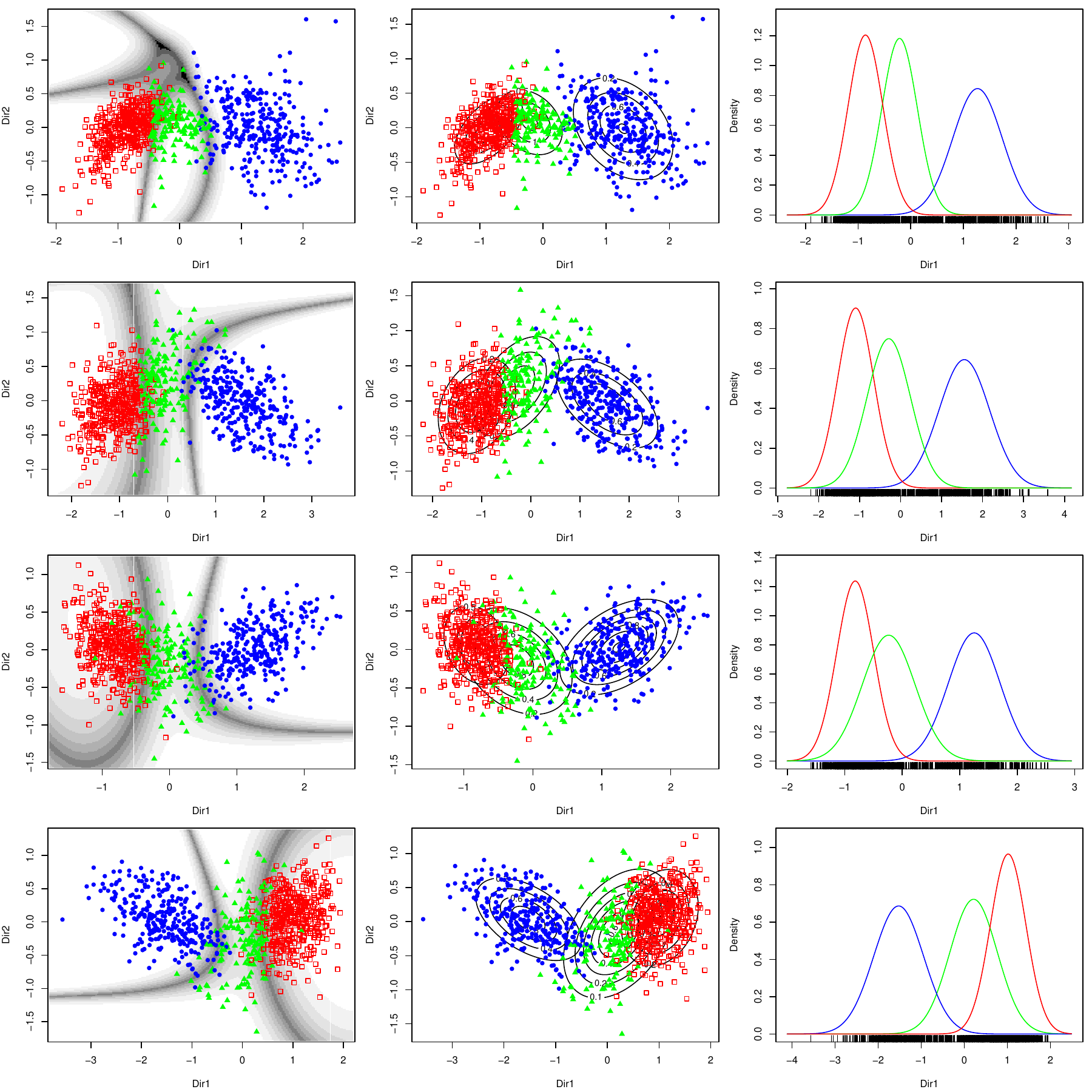}		
\caption{Dimension reduction for model-based clustering analysis (MclustDR),
the 3-dimensional data 
($\alpha_{\rm ph}$, $L_{\rm disk}$, and  $\lambda$=$L_{\rm disk}/L_{\rm Edd}$)(top No.1 row panels);
(CD, $L_{\rm disk}$, and  $\lambda$=$L_{\rm disk}/L_{\rm Edd}$)(top No.2 row panels);
and the 4-dimensional data 
($\Gamma_{\rm ph}$, CD, $L_{\rm disk}$, and  $\lambda$=$L_{\rm disk}/L_{\rm Edd}$)(top No.3 row panels);
(${\rm HR}_{45}$, CD, $L_{\rm disk}$, and  $\lambda$=$L_{\rm disk}/L_{\rm Edd}$)(lower No.4 row panels)
with
three groups (fsrqs, bl lacs, and CLBCs of the predictions) obtained from the $mclust$ clustering analysis.
The multidimensional data (see Figure \ref{Fig_combinations_3par},  and \ref{Fig_combinations_4par} respectively) are dimension reduction to two dimensions,
where the red hollow squares, and blue solid dots, and green triangles
indicate the predictions: fsrqs, bl lacs, and CLBCs respectively.
The grey colormap that represent the uncertainty boundaries of the mixture densities are presented in the left column (panels);
the scatter plots and contour plots are shown in the middle column (panels);
and the right column (panels) represents the density distributions of the predictions: fsrqs, bl lacs, and CLBCs respectively.
\label{Fig_DR_Gamma}}
\end{figure*}


\section{Results} \label{sec:Results}

Based on the BIC (e.g., see Table \ref{tab:parameter_subsets}), we find that there are 29 subsets that can estimate three candidate groups that are labeled as bl lacs, fsrqs, CLBCs. The three groups seem to correspond to three types in the sample that are BL Lacs, FSRQs, and CLBs.
Comparing the three groups  (bl lacs, fsrqs, and CLBCs) with three types (BL Lacs, FSRQs, CLBs), the ARIs are calculated for each combination of the 29 subsets.
We note that as the number of parameters increases, the ARI gradually reaches its maximum, where the ARI maximum is 0.636 with 4 parameters (see Table \ref{tab:parameter_subsets} and Figure \ref{Fig_ARI}). When 5 or more parameters are applied, the ARI starts to decline, which is similar to the results of our previous work (\citealt{2019ApJ...887..134K,2023MNRAS.525.3201K}).


\vspace{2.0mm}

In the 29 subsets with 3 groups, there are 4 subsets with the ARI $>$ 0.610 (see Table \ref{tab:parameter_subsets}), 
which are considered as the optimal parameters combinations (OPCs).
The plots of BIC using the mclust EVV model are shown in Figure \ref{Fig_BIC}. The BIC value with 3 components is biggest for the each 4 OPCs.
Where the No.68 subsets with 3 parameters ($\alpha_{\rm ph}$, CD,  and  $\lambda$=$L_{\rm disk}/L_{\rm Edd}$);
the No.89 subsets with 3 parameters (CD, $L_{\rm disk}$, and  $\lambda$=$L_{\rm disk}/L_{\rm Edd}$);
the No.124 subsets with 4 parameters  ($\Gamma_{\rm ph}$, CD, $L_{\rm disk}$, and  $\lambda$=$L_{\rm disk}/L_{\rm Edd}$);
and 
the No.158 subsets with 4 parameters (${\rm HR}_{45}$, CD, $L_{\rm disk}$, and  $\lambda$=$L_{\rm disk}/L_{\rm Edd}$),
which are marked as N* (e.g., No.68*, No.89*, No.124*, and No.158*) in the Column 1 of Table \ref{tab:parameter_subsets}.
The evaluation results of the No.68, No.89, No.124, and No.158 subsets are listed in Table \ref{tab:combinations_predictions}. 

\vspace{2.0mm}

The scatter plots of the results of clustering for No.68, No.89, No.124, and No.158 subsets are plotted in the upper right panels of Figure \ref{Fig_combinations_3par} and Figure \ref{Fig_combinations_4par} respectively, where the red hollow squares, and blue solid dots, and green triangles indicate the fsrq, bl lacs, and CLBCs respectively. 
In addition, the 2D distributions of each pairs of parameters for the three candidate sub-samples (fsrqs, bl lacs, and CLBCs) separately are fitted based on finite Gaussian mixture modelling. 
The Gaussian fitting results of each pairwise parameters are shown in the bottom left panels (the mirroring of upper right panels) of Figure \ref{Fig_combinations_3par} and \ref{Fig_combinations_4par} respectively,  where the best-fitting values and 1$\sigma$ confidence regions are marked as the crosses and the ellipse regions respectively.
We note that the group of CLB Candidates are obviously located between that of FSRQs and that of BL Lacs. Which are consistent with the results that the CLBs located between FSRQs and BL Lacs (\citealt{2024ApJ...962..122K}).

\vspace{2.0mm}

Using the dimension reduction function $mclustDR()$ to the clustering results of the each of the 4 subsets, a dimension reduction are performed to the multidimensional data with three groups that are dimension reduction to two dimensions. 
Where the red hollow squares, and blue solid dots, and green triangles indicate the predictions: fsrqs, bl lacs, and CLBCs respectively. The grey colormap that represent the uncertainty boundaries of the mixture densities are presented in the left column (panels); the scatter plots and contour plots are shown in the middle column (panels); and the right column (panels) represents the density distributions of the predictions: fsrqs, bl lacs, and CLBCs respectively.
In case of three groups in two dimensions, the CLB Candidates (green)  are also located between FSRQs  (red) and BL Lacs (blue) (see, Figure \ref{Fig_DR_Gamma}). Which are consistent with the results of our previous work (\citealt{2024ApJ...962..122K}).

\vspace{2.50mm}

There are 161, 156, 165, and 151 CLBCs that are predicted in No.68, No.89, No.124, and No.158 subsets respectively (see Table \ref{tab:parameter_subsets}, \ref{tab:combinations_predictions} or \ref{tab:selected}).
There are (total of) 217 possible CLBCs that are predicted in the 4 subsets and listed in Table \ref{tab:combinations_predictions}, where a source is predicted to be CLBCs by at least one subset, except for 64 CLBs reported in previous literature (see Table \ref{tab:selected}), there are 153 sources predicted as new possible CLB candidates.
Considering the evaluation results, the value of ARI (ARI $<$ 0.64) is a bit on the small side (see Table \ref{tab:parameter_subsets}), so we chose more subsets to construct our final analysis results.
The combined clustering results from the 4 subsets (cross-matching the clustering results of the 4 subsets) predict that there are 111 CLB candidates (see Table \ref{tab:selected}), including 44 CLBs reported in the previous literature.
The remaining 67 sources are new possible CLBCs, of which 11 CLBCs labeled as BL Lac and 56 CLBCs labeled as FSRQ in 4FGL-DR3 catalog.


\begin{deluxetable}{cccccccc}
\tablenum{3}
\tablecaption{Comparison of the sample with the prediction results of the selected subsets of parameters \label{tab:selected}}
\tablewidth{0pt}
\tablehead{
\colhead{Class} & \colhead{No.68} &  \colhead{No.89} & 
                              \colhead{No.124} & \colhead{No.158} &
                              \colhead{$N_{C4}$}  &\colhead{$N_{all}$}  &
                              \colhead{$N_{C2}$} 
                              }
\decimalcolnumbers
\startdata
$N_{p}$	&	161	&	156	&	165	&	151	&	111	&	217	  &  119  \\
bll 	&	17	&	14	&	21	&	21	&	11	&	26	  &   15  \\
CLB 	&	56	&	49	&	55	&	54	&	44	&	64	  &   46  \\
fsrq	&	88	&	93	&	89	&	76	&	56	&	127	  &   58  \\
\enddata
\tablecomments{
The classes of the sources in the sample are listed in Column 1. 
The results of comparison of No.68, No.89, No.124, and No.158 subsets (see Table \ref{tab:combinations_predictions}) and the sample are shown in  Columns 2-5 respectively.
The combined clustering results from the 4 subsets ($N_{C4}$) are shown in Columns 6.
The evaluation results of 4 subsets ($N_{all}$) are in Columns 7.
The combined clustering results from the
No.68 and No.158 subsets ($N_{C2}$) are shown in Columns 8.
 }
\end{deluxetable}



\section{discussion and conclusion} \label{sec:discussion_conclusion}

In the work, based on the 8 characteristic parameters of CLBs found in \cite{2024ApJ...962..122K}, using the $mclust$ Gaussian Mixture Modelling clustering algorithm, a clustering analysis is performed to search (evaluate) the CLB candidates. 
From the 255 subsets of the 8 characteristic parameters, we find that there are 29 subsets with 3 groups (corresponding to bl lacs, fsrqs, and CLBCs). Among of the 29 subsets, there are 4 subsets with ARI $>$ 0.610.
The combined clustering results from the 4 subsets predict that there are 111 CLB candidates, including 67 new probable CLBCs, of which 11 CLBCs labeled as BL Lac and 56 CLBCs labeled as FSRQ in 4FGL-DR3 catalog.

\vspace{2.0mm}

It should be noted that the projections for CLB candidates in this work are obviously different from those in \citealt{2022ApJ...935....4Z} and \citealt{2023MNRAS.525.3201K}.
In \cite{2022ApJ...935....4Z}, based on the analysis of the broad line region luminosity in Eddington units for a sample, they reported that there are 46 CLBCs including 38 BL Lacs, 7 FSRQs and 1 BCU (labeled in 4FGL-DR3 catalog), where, most of the predictions (38/46 $\simeq$ 82.60\%) are BL Lac type blazars.
However, in this present work, we suggest that there are 67 new probable CLBCs (11 BL Lacs, 56 FSRQs labeled in 4FGL-DR3 catalog), most of them (56/67 $\simeq$ 83.58\%) are FSRQ type blazars. Which is different with \cite{2022ApJ...935....4Z}'s predictions.
Only one source: 4FGL J1048.0$-$1912 (a LSP FSRQ) is evaluated as CLBC in both \cite{2022ApJ...935....4Z} and this work.
In \cite{2023MNRAS.525.3201K}, we suggested that there are 157 false LSP BL Lacs that are possible intrinsically FSRQs misclassified as BL Lacs are the most likely candidates of CLBs.
Among of 67 new probable CLBCs, there are 11 CLBCs labeled as BL Lac, of which 6 BL Lac type  CLBCs are labeled as LSP  BL Lacs. 
Only 3 of 6 BL Lacs (4FGL J0014.1$+$1910 4FGL J0105.1$+$3929 4FGL J2250.7$-$2806) are also suggested as CLBCs in  \cite{2023MNRAS.525.3201K}.

\begin{figure*}[htp!]
\centering
\includegraphics[width=18cm,height=12.3cm]{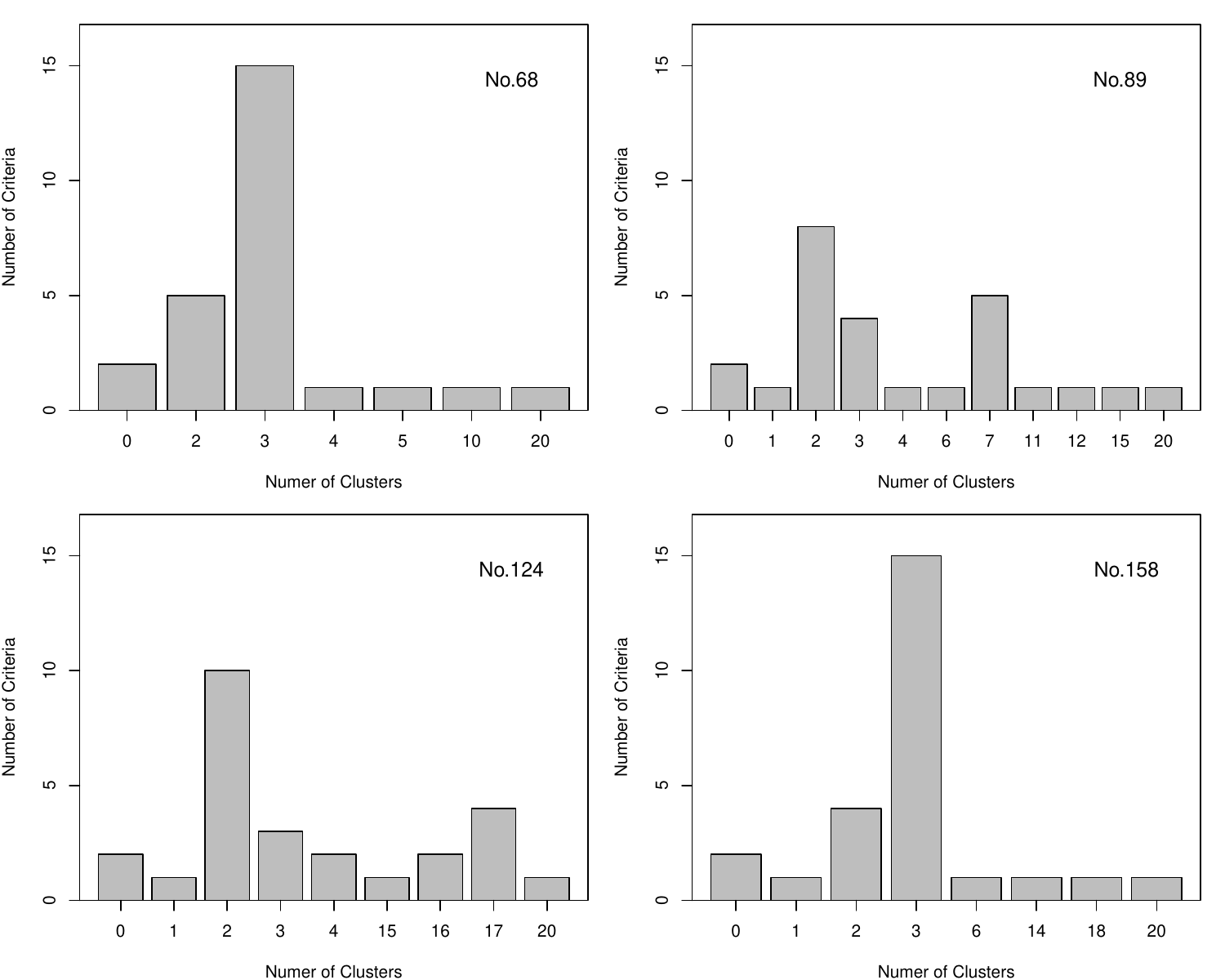}
\caption{The recommended number of clusters using 30 criteria provided by the $NbClust$ package for the 4 OPCs.
\label{Fig_NbClust}}
\end{figure*}

\vspace{2.0mm}

We also should be noted that the best number of mixing components (clusters) are selected only using the BIC. At the same time, the value of ARI $>$ 0.610 is also arbitrary to select OPCs. 
In order to check the the best number of clusters, $NbClust$ Package is also employed for determining the best number of clusters (\citealt{NbClust2014}) for the 4 OPCs.
Based on the criteria for determining the number of clusters is the largest, 
we found that the optimal clustering scheme for subsets of No.68 and No.158 (here, 15 criteria favor three clusters) is 3 clusters that is consistent with the results of $mclust$ clustering analysis.
However, the optimal clustering scheme for subsets of No.98 (8 criteria favor 2 clusters) and No.124 (10 criteria favor 2 clusters) is 2 clusters, which are different from the the number of 3 clusters of $mclust$ clustering analysis (see Figure \ref{Fig_NbClust}).
We note that the value of ARI $=$ 0.628 for the No.68 subsets and the value of ARI $=$ 0.636 for the No.158 subsets.
The value of ARI $>$ 0.610 that used to select OPCs should be arbitrary.
If the value of ARI $>$ 0.626 is to select the OPCs.
There are only two subsets of No.68 and No.158 will be selected as the OPCs. The best number of mixing components (clusters) are agree in 
the subsets of No.68 and No.158 parameters.
If we chose the two subsets (No.68 and No.158) to construct our analysis results.
The combined clustering results from the two subsets (No.68 and No.158) will predict 119 CLB candidates that including 46 CLBs (see Table \ref{tab:selected}).
The other 73 sources are new possible CLBCs, of which 15 CLBCs labeled as BL Lac and 58 CLBCs labeled as FSRQ in 4FGL-DR3 catalog.

\begin{figure*}[htp!]
\centering
\includegraphics[width=16cm,height=10cm]{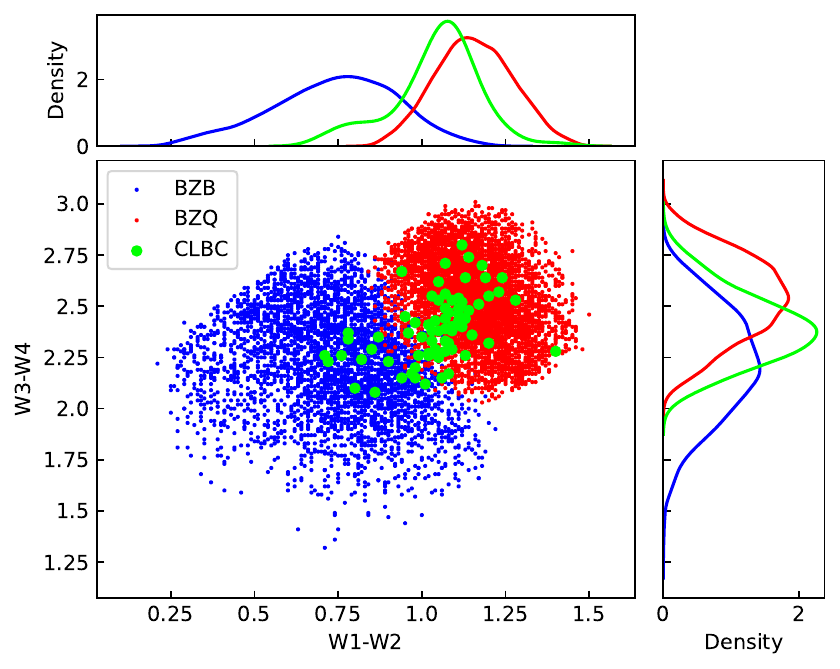}
\caption{74 CLBCs and WISE Blazars' $W1-W2$ vs. $W3-W4$ color-color diagram. The red and blue solid dots and lines respectively represent the BZQ 
(corresponding to FSRQs) and BZB (corresponding to BL Lacs) obtained from \cite{2019ApJS..242....4D}. The green solid dots and lines represent the 74 CLBCs predicted in the work. The $W1-W2$ and $W3-W4$ data used are obtained from \cite{2019ApJS..242....4D}.
\label{wise}}
\end{figure*}

\vspace{2.0mm}

Cross-matching the 111 CLBCs with the 9541 WISE Blazars obtained from \cite{2019ApJS..242....4D}, the color data of 74 CLBCs are obtained. The color-color diagram of scatter ($W1-W2$ vs. $W3-W4$ ) and density distribution are shown in Figure~\ref{wise}.
We note that the group of CLBCs are also located between that of BZQ (corresponding to FSRQs) and that of BZB (corresponding to BL Lacs) in WISE Blazar color-color diagram. It is consistent with our previous research results (\citealt{2024ApJ...962..122K}).
The physical properties of CLBs, as well as their potential similarities in other parameter spaces, necessitate further address in future studies.

\vspace{2.0mm}

In addition, we also should note that, in this work, our results are obtained only by $mclust$ clustering analysis, based on the data obtained from the 8 physical properties parameters (\citealt{2024ApJ...962..122K}), adding no external data.
Selection effects of sample and clustering methods may affect the source distributions and affect the results of the clustering analysis in this work. 

Although the clustering results may be subject to selection effects, such as a relatively small sample size, it is important to note that the data completeness is limited, with only 925 blazars having measurements of CD, $L_{\rm disk}$, and $\lambda = L_{\rm disk}/L_{\rm Edd}$  \citep{2021ApJS..253...46P}
and the $mclust$ clustering algorithm may not be optimal, it may be helpful for source selections in searching, evaluation and certification of CLBs in the future or to providing some clues for future studying of the physical properties of CLBs.
Other more preferable clustering methods that uses a large and more complete sample is considered and needed to further test and address the issue.



\begin{acknowledgments}

We thank the editor and anonymous referee for very constructive and helpful comments and suggestions, which greatly helped us to improve our paper. This work is partially supported by the National Natural Science Foundation of China (grant Nos. 12163002, U1931203, U2031201, and 12363002, 12233007), the National SKA Program of China (grant No. 2022SKA0120101), and the Discipline Team of Liupanshui Normal University (grant No. LPSSY2023XKTD11).

\end{acknowledgments}

\vspace{5mm}
\facilities{Fermi(LAT)}

\software{R \citep{R_code},  mclust \citep{mclust_2016,mclust_book}, NbClust \citep{NbClust2014}.  }

\bibliography{aastex631}{}

\begin{thebibliography}{}
\expandafter\ifx\csname natexlab\endcsname\relax\def\natexlab#1{#1}\fi
\providecommand{\url}[1]{\href{#1}{#1}}
\providecommand{\dodoi}[1]{doi:~\href{http://doi.org/#1}{\nolinkurl{#1}}}
\providecommand{\doeprint}[1]{\href{http://ascl.net/#1}{\nolinkurl{http://ascl.net/#1}}}
\providecommand{\doarXiv}[1]{\href{https://arxiv.org/abs/#1}{\nolinkurl{https://arxiv.org/abs/#1}}}

\bibitem[{{Abdollahi} {et~al.}(2022){Abdollahi}, {Acero}, {Baldini}, {Ballet}, {Bastieri}, {Bellazzini}, {Berenji}, {Berretta}, {Bissaldi}, {Blandford}, {Bloom}, {Bonino}, {Brill}, {Britto}, {Bruel}, {Burnett}, {Buson}, {Cameron}, {Caputo}, {Caraveo}, {Castro}, {Chaty}, {Cheung}, {Chiaro}, {Cibrario}, {Ciprini}, {Coronado-Bl{\'a}zquez}, {Crnogorcevic}, {Cutini}, {D'Ammando}, {De Gaetano}, {Digel}, {Di Lalla}, {Dirirsa}, {Di Venere}, {Dom{\'\i}nguez}, {Fallah Ramazani}, {Fegan}, {Ferrara}, {Fiori}, {Fleischhack}, {Franckowiak}, {Fukazawa}, {Funk}, {Fusco}, {Galanti}, {Gammaldi}, {Gargano}, {Garrappa}, {Gasparrini}, {Giacchino}, {Giglietto}, {Giordano}, {Giroletti}, {Glanzman}, {Green}, {Grenier}, {Grondin}, {Guillemot}, {Guiriec}, {Gustafsson}, {Harding}, {Hays}, {Hewitt}, {Horan}, {Hou}, {J{\'o}hannesson}, {Karwin}, {Kayanoki}, {Kerr}, {Kuss}, {Landriu}, {Larsson}, {Latronico}, {Lemoine-Goumard}, {Li}, {Liodakis}, {Longo}, {Loparco}, {Lott}, {Lubrano}, {Maldera}, {Malyshev}, {Manfreda}, {Mart{\'\i}-Devesa},
  {Mazziotta}, {Mereu}, {Meyer}, {Michelson}, {Mirabal}, {Mitthumsiri}, {Mizuno}, {Moiseev}, {Monzani}, {Morselli}, {Moskalenko}, {Negro}, {Nuss}, {Omodei}, {Orienti}, {Orlando}, {Paneque}, {Pei}, {Perkins}, {Persic}, {Pesce-Rollins}, {Petrosian}, {Pillera}, {Poon}, {Porter}, {Principe}, {Rain{\`o}}, {Rando}, {Rani}, {Razzano}, {Razzaque}, {Reimer}, {Reimer}, {Reposeur}, {S{\'a}nchez-Conde}, {Saz Parkinson}, {Scotton}, {Serini}, {Sgr{\`o}}, {Siskind}, {Smith}, {Spandre}, {Spinelli}, {Sueoka}, {Suson}, {Tajima}, {Tak}, {Thayer}, {Thompson}, {Torres}, {Troja}, {Valverde}, {Wood}, \& {Zaharijas}}]{2022ApJS..260...53A}
{Abdollahi}, S., {Acero}, F., {Baldini}, L., {et~al.} 2022, \apjs, 260, 53, \dodoi{10.3847/1538-4365/ac6751}

\bibitem[{{Ackermann} {et~al.}(2012){Ackermann}, {Ajello}, {Allafort}, {Antolini}, {Baldini}, {Ballet}, {Barbiellini}, {Bastieri}, {Bellazzini}, {Berenji}, {Blandford}, {Bloom}, {Bonamente}, {Borgland}, {Bouvier}, {Brandt}, {Bregeon}, {Brigida}, {Bruel}, {Buehler}, {Burnett}, {Buson}, {Caliandro}, {Cameron}, {Caraveo}, {Casandjian}, {Cavazzuti}, {Cecchi}, {{\c{C}}elik}, {Charles}, {Chekhtman}, {Chen}, {Cheung}, {Chiang}, {Ciprini}, {Claus}, {Cohen-Tanugi}, {Conrad}, {Cutini}, {de Angelis}, {DeCesar}, {De Luca}, {de Palma}, {Dermer}, {Silva}, {Drell}, {Drlica-Wagner}, {Dubois}, {Enoto}, {Favuzzi}, {Fegan}, {Ferrara}, {Focke}, {Fortin}, {Fukazawa}, {Funk}, {Fusco}, {Gargano}, {Gasparrini}, {Gehrels}, {Germani}, {Giglietto}, {Giordano}, {Giroletti}, {Glanzman}, {Godfrey}, {Grenier}, {Grondin}, {Grove}, {Guillemot}, {Guiriec}, {Gustafsson}, {Hadasch}, {Hanabata}, {Harding}, {Hayashida}, {Hays}, {Healey}, {Hill}, {Horan}, {Hou}, {J{\'o}hannesson}, {Johnson}, {Johnson}, {Kamae}, {Katagiri}, {Kataoka}, {Kerr},
  {Kn{\"o}dlseder}, {Kuss}, {Lande}, {Latronico}, {Lee}, {Lemoine-Goumard}, {Longo}, {Loparco}, {Lott}, {Lovellette}, {Lubrano}, {Madejski}, {Mazziotta}, {McEnery}, {Mehault}, {Michelson}, {Mignani}, {Mitthumsiri}, {Mizuno}, {Monte}, {Monzani}, {Morselli}, {Moskalenko}, {Murgia}, {Nakamori}, {Naumann-Godo}, {Nolan}, {Norris}, {Nuss}, {Ohsugi}, {Okumura}, {Omodei}, {Orlando}, {Ormes}, {Ozaki}, {Paneque}, {Panetta}, {Parent}, {Pelassa}, {Pesce-Rollins}, {Pierbattista}, {Piron}, {Pivato}, {Porter}, {Rain{\`o}}, {Rando}, {Ray}, {Razzano}, {Reimer}, {Reimer}, {Reposeur}, {Romani}, {Sadrozinski}, {Salvetti}, {Saz Parkinson}, {Schalk}, {Sgr{\`o}}, {Shaw}, {Siskind}, {Smith}, {Spandre}, {Spinelli}, {Suson}, {Takahashi}, {Tanaka}, {Thayer}, {Thayer}, {Thompson}, {Tibaldo}, {Tibolla}, {Torres}, {Tosti}, {Tramacere}, {Troja}, {Usher}, {Vandenbroucke}, {Vasileiou}, {Vianello}, {Vilchez}, {Vitale}, {Waite}, {Wallace}, {Wang}, {Winer}, {Wolff}, {Wood}, {Wood}, {Yang}, \& {Zimmer}}]{2012ApJ...753...83A}
{Ackermann}, M., {Ajello}, M., {Allafort}, A., {et~al.} 2012, \apj, 753, 83, \dodoi{10.1088/0004-637X/753/1/83}

\bibitem[{{Ajello} {et~al.}(2022){Ajello}, {Baldini}, {Ballet}, {Bastieri}, {Becerra Gonzalez}, {Bellazzini}, {Berretta}, {Bissaldi}, {Bonino}, {Brill}, {Bruel}, {Buson}, {Caputo}, {Caraveo}, {Cheung}, {Chiaro}, {Cibrario}, {Ciprini}, {Crnogorcevic}, {Cutini}, {D'Ammando}, {De Gaetano}, {Di Lalla}, {Di Venere}, {Dom{\'\i}nguez}, {Ramazani}, {Ferrara}, {Fiori}, {Fukazawa}, {Funk}, {Fusco}, {Gammaldi}, {Gargano}, {Garrappa}, {Gasparrini}, {Giglietto}, {Giordano}, {Giroletti}, {Green}, {Grenier}, {Guiriec}, {Horan}, {Hou}, {Kayanoki}, {Kuss}, {Larsson}, {Latronico}, {Lewis}, {Li}, {Liodakis}, {Longo}, {Loparco}, {Lott}, {Lovellette}, {Lubrano}, {Madejski}, {Maldera}, {Manfreda}, {Mart{\'\i}-Devesa}, {Mazziotta}, {Mereu}, {Michelson}, {Mirabal}, {Mitthumsiri}, {Mizuno}, {Monzani}, {Morselli}, {Moskalenko}, {Negro}, {Ojha}, {Orienti}, {Orlando}, {Ormes}, {Pei}, {Pe{\~n}a-Herazo}, {Persic}, {Pesce-Rollins}, {Petrosian}, {Pillera}, {Poon}, {Porter}, {Principe}, {Rain{\`o}}, {Rando}, {Rani}, {Razzano}, {Razzaque},
  {Reimer}, {Reimer}, {Scotton}, {Serini}, {Sgr{\`o}}, {Siskind}, {Spandre}, {Spinelli}, {Suson}, {Tajima}, {Torres}, {Valverde}, {Yassin}, \& {Zaharijas}}]{2022ApJS..263...24A}
{Ajello}, M., {Baldini}, L., {Ballet}, J., {et~al.} 2022, \apjs, 263, 24, \dodoi{10.3847/1538-4365/ac9523}

\bibitem[{{{\'A}lvarez Crespo} {et~al.}(2016){{\'A}lvarez Crespo}, {Masetti}, {Ricci}, {Landoni}, {Pati{\~n}o-{\'A}lvarez}, {Massaro}, {D'Abrusco}, {Paggi}, {Chavushyan}, {Jim{\'e}nez-Bail{\'o}n}, {Torrealba}, {Latronico}, {La Franca}, {Smith}, \& {Tosti}}]{2016AJ....151...32A}
{{\'A}lvarez Crespo}, N., {Masetti}, N., {Ricci}, F., {et~al.} 2016, \aj, 151, 32, \dodoi{10.3847/0004-6256/151/2/32}

\bibitem[{{Baron}(2019)}]{2019arXiv190407248B}
{Baron}, D. 2019, arXiv e-prints, arXiv:1904.07248, \dodoi{10.48550/arXiv.1904.07248}

\bibitem[{{Bianchin} {et~al.}(2009){Bianchin}, {Foschini}, {Ghisellini}, {Tagliaferri}, {Tavecchio}, {Treves}, {Di Cocco}, {Gliozzi}, {Pian}, {Sambruna}, \& {Wolter}}]{2009A&A...496..423B}
{Bianchin}, V., {Foschini}, L., {Ghisellini}, G., {et~al.} 2009, \aap, 496, 423, \dodoi{10.1051/0004-6361/200811128}

\bibitem[{Charrad {et~al.}(2014)Charrad, Ghazzali, Boiteau, \& Niknafs}]{NbClust2014}
Charrad, M., Ghazzali, N., Boiteau, V., \& Niknafs, A. 2014, Journal of Statistical Software, 61, 1.
\newblock \url{https://www.jstatsoft.org/v61/i06/}

\bibitem[{{D'Abrusco} {et~al.}(2019){D'Abrusco}, {{\'A}lvarez Crespo}, {Massaro}, {Campana}, {Chavushyan}, {Landoni}, {La Franca}, {Masetti}, {Milisavljevic}, {Paggi}, {Ricci}, \& {Smith}}]{2019ApJS..242....4D}
{D'Abrusco}, R., {{\'A}lvarez Crespo}, N., {Massaro}, F., {et~al.} 2019, \apjs, 242, 4, \dodoi{10.3847/1538-4365/ab16f4}

\bibitem[{{D'Elia} {et~al.}(2015){D'Elia}, {Padovani}, {Giommi}, \& {Turriziani}}]{2015MNRAS.449.3517D}
{D'Elia}, V., {Padovani}, P., {Giommi}, P., \& {Turriziani}, S. 2015, \mnras, 449, 3517, \dodoi{10.1093/mnras/stv573}

\bibitem[{{Foschini} {et~al.}(2021){Foschini}, {Lister}, {Ant{\'o}n}, {Berton}, {Ciroi}, {March{\~a}}, {Tornikoski}, {J{\"a}rvel{\"a}}, {Romano}, {Vercellone}, \& {Dalla Bont{\`a}}}]{2021Univ....7..372F}
{Foschini}, L., {Lister}, M.~L., {Ant{\'o}n}, S., {et~al.} 2021, Universe, 7, 372, \dodoi{10.3390/universe7100372}

\bibitem[{{Foschini} {et~al.}(2022){Foschini}, {Lister}, {Andernach}, {Ciroi}, {Marziani}, {Ant{\'o}n}, {Berton}, {Dalla Bont{\`a}}, {J{\"a}rvel{\"a}}, {March{\~a}}, {Romano}, {Tornikoski}, {Vercellone}, \& {Vietri}}]{2022Univ....8..587F}
{Foschini}, L., {Lister}, M.~L., {Andernach}, H., {et~al.} 2022, Universe, 8, 587, \dodoi{10.3390/universe8110587}

\bibitem[{{Ghisellini} {et~al.}(2012){Ghisellini}, {Tavecchio}, {Foschini}, {Sbarrato}, {Ghirlanda}, \& {Maraschi}}]{2012MNRAS.425.1371G}
{Ghisellini}, G., {Tavecchio}, F., {Foschini}, L., {et~al.} 2012, \mnras, 425, 1371, \dodoi{10.1111/j.1365-2966.2012.21554.x}

\bibitem[{{Giommi} {et~al.}(2012){Giommi}, {Padovani}, {Polenta}, {Turriziani}, {D'Elia}, \& {Piranomonte}}]{2012MNRAS.420.2899G}
{Giommi}, P., {Padovani}, P., {Polenta}, G., {et~al.} 2012, \mnras, 420, 2899, \dodoi{10.1111/j.1365-2966.2011.20044.x}

\bibitem[{Hubert \& Arabie(1985)}]{Hubert_Arabie_1985}
Hubert, L., \& Arabie, P. 1985, Journal of Classification, 2, 193, \dodoi{10.1007/BF01908075}

\bibitem[{{Hutsem{\'e}kers} {et~al.}(2019){Hutsem{\'e}kers}, {Ag{\'\i}s Gonz{\'a}lez}, {Marin}, {Sluse}, {Ramos Almeida}, \& {Acosta Pulido}}]{2019A&A...625A..54H}
{Hutsem{\'e}kers}, D., {Ag{\'\i}s Gonz{\'a}lez}, B., {Marin}, F., {et~al.} 2019, \aap, 625, A54, \dodoi{10.1051/0004-6361/201834633}

\bibitem[{Kang(2023)}]{shi_ju_kang_2023_10061349}
Kang, S.-J. 2023, {TCLBCat: A Changing-Look (Transition) Blazars Catalog} (Zenodo), \dodoi{10.5281/zenodo.10061349}

\bibitem[{{Kang} {et~al.}(2019){Kang}, {Li}, {Ou}, {Zhu}, {Fan}, {Wu}, \& {Yin}}]{2019ApJ...887..134K}
{Kang}, S.-J., {Li}, E., {Ou}, W., {et~al.} 2019, \apj, 887, 134, \dodoi{10.3847/1538-4357/ab558b}

\bibitem[{{Kang} {et~al.}(2024){Kang}, {Lyu}, {Wu}, {Zheng}, \& {Fan}}]{2024ApJ...962..122K}
{Kang}, S.-J., {Lyu}, B., {Wu}, Q., {Zheng}, Y.-G., \& {Fan}, J. 2024, \apj, 962, 122, \dodoi{10.3847/1538-4357/ad0fdf}

\bibitem[{{Kang} {et~al.}(2023){Kang}, {Zheng}, \& {Wu}}]{2023MNRAS.525.3201K}
{Kang}, S.-J., {Zheng}, Y.-G., \& {Wu}, Q. 2023, \mnras, 525, 3201, \dodoi{10.1093/mnras/stad2456}

\bibitem[{{Mishra}(2021)}]{2021AAS...23740807M}
{Mishra}, H. 2021, in American Astronomical Society Meeting Abstracts, Vol.~53, American Astronomical Society Meeting Abstracts, 408.07

\bibitem[{{Mishra} {et~al.}(2021){Mishra}, {Dai}, {Chen}, {Cheng}, {Jayasinghe}, {Tucker}, {Vallely}, {Bersier}, {Bose}, {Do}, {Dong}, {Holoien}, {Huber}, {Kochanek}, {Liang}, {Payne}, {Prieto}, {Shappee}, {Stanek}, {Bhatiani}, {Cox}, {DeFrancesco}, {Shen}, {Thompson}, \& {Wang}}]{2021ApJ...913..146M}
{Mishra}, H.~D., {Dai}, X., {Chen}, P., {et~al.} 2021, \apj, 913, 146, \dodoi{10.3847/1538-4357/abf63d}

\bibitem[{{Paiano} {et~al.}(2024){Paiano}, {Falomo}, {Treves}, {Scarpa}, \& {Sbarufatti}}]{2024ApJ...968...81P}
{Paiano}, S., {Falomo}, R., {Treves}, A., {Scarpa}, R., \& {Sbarufatti}, B. 2024, \apj, 968, 81, \dodoi{10.3847/1538-4357/ad4a56}

\bibitem[{{Paliya} {et~al.}(2021){Paliya}, {Dom{\'\i}nguez}, {Ajello}, {Olmo-Garc{\'\i}a}, \& {Hartmann}}]{2021ApJS..253...46P}
{Paliya}, V.~S., {Dom{\'\i}nguez}, A., {Ajello}, M., {Olmo-Garc{\'\i}a}, A., \& {Hartmann}, D. 2021, \apjs, 253, 46, \dodoi{10.3847/1538-4365/abe135}

\bibitem[{{Pandey} {et~al.}(2023){Pandey}, {Kushwaha}, {Wiita}, {Prince}, {Czerny}, \& {Stalin}}]{2023arXiv231005096P}
{Pandey}, A., {Kushwaha}, P., {Wiita}, P.~J., {et~al.} 2023, arXiv e-prints, arXiv:2310.05096, \dodoi{10.48550/arXiv.2310.05096}

\bibitem[{{Pasham} \& {Wevers}(2019)}]{2019RNAAS...3...92P}
{Pasham}, D.~R., \& {Wevers}, T. 2019, Research Notes of the American Astronomical Society, 3, 92, \dodoi{10.3847/2515-5172/ab304a}

\bibitem[{{Pe{\~n}a-Herazo} {et~al.}(2021){Pe{\~n}a-Herazo}, {Massaro}, {Gu}, {Paggi}, {Landoni}, {D'Abrusco}, {Ricci}, {Masetti}, \& {Chavushyan}}]{2021AJ....161..196P}
{Pe{\~n}a-Herazo}, H.~A., {Massaro}, F., {Gu}, M., {et~al.} 2021, \aj, 161, 196, \dodoi{10.3847/1538-3881/abe41d}

\bibitem[{{R Core Team}(2022)}]{R_code}
{R Core Team}. 2022, R: A Language and Environment for Statistical Computing, R Foundation for Statistical Computing, Vienna, Austria.
\newblock \url{https://www.R-project.org/}

\bibitem[{{Ren} {et~al.}(2024){Ren}, {Zhou}, {Zheng}, {Kang}, \& {Wu}}]{2024arXiv240217099R}
{Ren}, S.~S., {Zhou}, R.~X., {Zheng}, Y.~G., {Kang}, S.~J., \& {Wu}, Q. 2024, arXiv e-prints, arXiv:2402.17099, \dodoi{10.48550/arXiv.2402.17099}

\bibitem[{{Ricci} \& {Trakhtenbrot}(2022)}]{2022arXiv221105132R}
{Ricci}, C., \& {Trakhtenbrot}, B. 2022, arXiv e-prints, arXiv:2211.05132, \dodoi{10.48550/arXiv.2211.05132}

\bibitem[{{Ruan} {et~al.}(2014){Ruan}, {Anderson}, {Plotkin}, {Brandt}, {Burnett}, {Myers}, \& {Schneider}}]{2014ApJ...797...19R}
{Ruan}, J.~J., {Anderson}, S.~F., {Plotkin}, R.~M., {et~al.} 2014, \apj, 797, 19, \dodoi{10.1088/0004-637X/797/1/19}

\bibitem[{Scrucca {et~al.}(2016)Scrucca, Fop, Murphy, \& Raftery}]{mclust_2016}
Scrucca, L., Fop, M., Murphy, T.~B., \& Raftery, A.~E. 2016, The {R} Journal, 8, 289.
\newblock \url{https://doi.org/10.32614/RJ-2016-021}

\bibitem[{Scrucca {et~al.}(2023)Scrucca, Fraley, Murphy, \& Raftery}]{mclust_book}
Scrucca, L., Fraley, C., Murphy, T.~B., \& Raftery, A.~E. 2023, Model-Based Clustering, Classification, and Density Estimation Using {mclust} in {R} (Chapman and Hall/CRC), \dodoi{10.1201/9781003277965}

\bibitem[{{Stickel} {et~al.}(1991){Stickel}, {Padovani}, {Urry}, {Fried}, \& {Kuehr}}]{1991ApJ...374..431S}
{Stickel}, M., {Padovani}, P., {Urry}, C.~M., {Fried}, J.~W., \& {Kuehr}, H. 1991, \apj, 374, 431, \dodoi{10.1086/170133}

\bibitem[{{Stocke} {et~al.}(1991){Stocke}, {Morris}, {Gioia}, {Maccacaro}, {Schild}, {Wolter}, {Fleming}, \& {Henry}}]{1991ApJS...76..813S}
{Stocke}, J.~T., {Morris}, S.~L., {Gioia}, I.~M., {et~al.} 1991, \apjs, 76, 813, \dodoi{10.1086/191582}

\bibitem[{{Urry} \& {Padovani}(1995)}]{1995PASP..107..803U}
{Urry}, C.~M., \& {Padovani}, P. 1995, \pasp, 107, 803, \dodoi{10.1086/133630}

\bibitem[{{Vermeulen} {et~al.}(1995){Vermeulen}, {Ogle}, {Tran}, {Browne}, {Cohen}, {Readhead}, {Taylor}, \& {Goodrich}}]{1995ApJ...452L...5V}
{Vermeulen}, R.~C., {Ogle}, P.~M., {Tran}, H.~D., {et~al.} 1995, \apjl, 452, L5, \dodoi{10.1086/309716}

\bibitem[{{Xiao} {et~al.}(2022){Xiao}, {Fan}, {Ouyang}, {Hu}, {Chen}, {Fu}, \& {Zhang}}]{2022ApJ...936..146X}
{Xiao}, H., {Fan}, J., {Ouyang}, Z., {et~al.} 2022, \apj, 936, 146, \dodoi{10.3847/1538-4357/ac887f}

\bibitem[{{Zhang} {et~al.}(2022){Zhang}, {Liu}, \& {Fan}}]{2022ApJ...935....4Z}
{Zhang}, L., {Liu}, Y., \& {Fan}, J. 2022, \apj, 935, 4, \dodoi{10.3847/1538-4357/ac7bde}

\end{thebibliography}
\bibliographystyle{aasjournal}

\end{CJK*}
\end{document}